\documentclass[aps,prd,superscriptaddress,A4paper,preprintnumbers,nofootinbib]{revtex4}
\usepackage{amssymb,amsmath,graphicx,epsfig,bm,color}
\usepackage{bm}
 \usepackage{float}
\usepackage{slashed}
\usepackage{verbatim}
\newcommand{\be}{\begin{eqnarray}}
\newcommand{\ee}{\end{eqnarray}}
\usepackage[colorlinks]{hyperref}
\usepackage{leftidx}
\hypersetup{
    colorlinks=red,
    citecolor=blue,
    filecolor=magenta,
    urlcolor=red,
}
\renewcommand{\d}{\mathrm{d}}
\newcommand{\pup}{p^\uparrow}

\newcommand{\bfk}{\mbox{\boldmath $k$}}
\begin{document}

\title{Single-spin asymmetries in $p^\uparrow p \to J/\psi +X$ within a TMD approach: \\
role of the color octet mechanism
}

\author{Umberto D'Alesio}
\email{umberto.dalesio@ca.infn.it}
\affiliation{Dipartimento di Fisica, Universit\`a di Cagliari, Cittadella Universitaria, I-09042 Monserrato (CA), Italy}
\affiliation{INFN, Sezione di Cagliari, Cittadella Universitaria, I-09042 Monserrato (CA), Italy}

\author{Francesco Murgia}
\email{francesco.murgia@ca.infn.it}
\affiliation{INFN, Sezione di Cagliari, Cittadella Universitaria, I-09042 Monserrato (CA), Italy}

\author{Cristian Pisano}
\email{cristian.pisano@ca.infn.it}
\affiliation{Dipartimento di Fisica, Universit\`a di Cagliari, Cittadella Universitaria, I-09042 Monserrato (CA), Italy}
\affiliation{INFN, Sezione di Cagliari, Cittadella Universitaria, I-09042 Monserrato (CA), Italy}

\author{Sangem Rajesh}
\email{rajesh.sangem@ca.infn.it}
\affiliation{Dipartimento di Fisica, Universit\`a di Cagliari, Cittadella Universitaria, I-09042 Monserrato (CA), Italy}
\affiliation{INFN, Sezione di Cagliari, Cittadella Universitaria, I-09042 Monserrato (CA), Italy}

\date{\today}
\begin{abstract}
We calculate the transverse single-spin asymmetry (SSA) for $J/\psi$ production in proton-proton collisions, within non-relativistic QCD, employing the transverse momentum dependent, generalized parton model, which includes both spin and intrinsic motion effects. In particular, we study the role of the color octet mechanism in accessing the gluon Sivers function.
In doing that, we also show, within this approach, how the singularities coming from color octet terms in the low-$P_T$ region can be handled, leading to finite cross sections.
Predictions for both unpolarized cross sections and SSAs are given and compared against PHENIX data. Estimates for LHCb in the fixed target mode are also presented.
\end{abstract}
\maketitle

\section{Introduction}
\label{sec1}

Transverse momentum dependent parton distribution (TMD-PDFs) and fragmentation functions (TMD-FFs), shortly referred to as TMDs, have been receiving significant interest, both theoretically and experimentally, for their important role in mapping the three-dimensional structure of the proton in momentum space and as a tool to explain several observed azimuthal and single-spin asymmetries (SSAs).
Among the eight leading-twist TMD-PDFs, the Sivers function~\cite{Sivers:1989cc,Sivers:1990fh} is certainly the most studied for its very interesting properties.
It describes the asymmetric azimuthal distribution of unpolarized quarks or gluons in a transversely polarized, highly energetic proton, encoding correlations between the transverse momentum of the parton and the spin of the proton. As a result, it could lead to transverse single-spin asymmetries in scattering processes where one of the initial hadrons is transversely polarized. It could also play a role in our understanding of the proton spin in terms of the spin and orbital angular momentum  of its constituents.
Moreover, the Sivers function as probed in semi-inclusive deep inelastic scattering (SIDIS) processes is expected to have an opposite sign when probed in Drell-Yan (DY) processes~\cite{Collins:2002kn}. This type of non-universal property of the Sivers function could be in principle tested experimentally and represents a key issue for our understanding of SSAs in QCD.

While for the quark Sivers function the amount of information is certainly well consolidated, thanks to several experimental results and their phenomenological analyses, very little is known on its gluon counterpart~\cite{Boer:2015vso}.
SSAs have been observed in processes like SIDIS~\cite{Airapetian:2004tw,Adolph:2012sp} and inclusive single-hadron production in $pp$ collisions~\cite{Adare:2013ekj,Aidala:2018gmp}, see also Refs.~\cite{DAlesio:2007bjf,Barone:2010zz,Aschenauer:2015ndk} for recent reviews.
The first class of processes, characterized by two well separate energy scales, are usually analyzed within a TMD factorization scheme, nowadays well established from the theoretical point of view~\cite{Ji:2004xq,Collins:2011zzd,GarciaEchevarria:2011rb}. For single inclusive hadron production, where only one large scale is present, the Generalized Parton Model (GPM) approach has been developed (see Ref.~\cite{DAlesio:2007bjf} and references therein). This extends the usual collinear parton model, including both spin and transverse momentum effects, and is formulated on a phenomenological basis. Nevertheless, it has been shown to be able to describe many polarization effects and it is therefore worth to be pursued. For the sake of completeness, we have to mention that an alternative collinear formalism, the twist-3 approach, for which factorization has been proven, has been also successfully applied for this class of processes~\cite{Qiu:1998ia,Kouvaris:2006zy,Kanazawa:2014dca}.

Within the GPM, some attempts to extract the gluon Sivers function (GSF) have been made in Refs.~\cite{DAlesio:2015fwo,DAlesio:2018rnv}, by fitting mid-rapidity pion SSA data at RHIC~\cite{Adare:2013ekj}. A similar analysis has been performed in Ref.~\cite{Godbole:2017syo}.
In Refs.~\cite{DAlesio:2017rzj,DAlesio:2018rnv}, $J/\psi$ production in $pp$ collisions was considered, within the color singlet model, as a tool to learn on the GSF. The advantage of this process is that it allows to directly probe gluon TMDs due to the negligible contribution from quark induced subprocesses. The effects of initial- and final-state interactions on the Sivers function were also considered by adopting, in the one-gluon-exchange approximation, the so-called color-gauge invariant version of the GPM (CGI-GPM). In particular, in Ref.~\cite{DAlesio:2018rnv}, from a combined analysis of mid-rapidity pion and $D$-meson production data, the first constraints on the two independent gluon Sivers functions, emerging in this approach, were given. It was shown that using the so extracted GSFs allows for a fairly good description of PHENIX SSA data~\cite{Aidala:2018gmp} for $J/\psi$ production, very similarly to what obtained in the GPM. Complementary studies to access the GSF in $ep$ collisions were performed in Refs.~\cite{Godbole:2012bx,Godbole:2014tha,Mukherjee:2016qxa,Boer:2016fqd,Rajesh:2018qks,Bacchetta:2018ivt,Zheng:2018awe}.

It is important to recall, at this stage, that the heavy quarkonium production mechanism is not yet completely understood and several approaches have been proposed. Among the first ones exploited in the literature, we mention: the color singlet model (CSM)~\cite{Carlson:1976cd} and the color evaporation model (CEM)~\cite{Halzen:1977rs}. A more formal scheme was then developed, based on a detailed separation of the involved energy scales: the non-relativistic QCD (NRQCD) approach, an  effective field theory which provides a rigourous treatment of heavy-quarkonium production and decay~\cite{Bodwin:1994jh}.
The common feature of these approaches is that quarkonium production is factorized into short-distance and
long-distance parts. The short-distance piece, i.e.~the formation of the heavy quark-antiquark pair, can be calculated using perturbative QCD, while the long-distance part contains the nonperturbative information related to the hadronization of the heavy-quark pair into a physical quarkonium state: this is encoded in the so-called long-distance matrix elements (LDMEs). These have to be extracted from fits to experimental data and, even if in principle should be universal, there are very different sets in the literature, depending on theoretical as well as phenomenological assumptions.

In the CSM, the heavy-quark pair is produced directly in a color singlet (CS) state with the quantum numbers as the observed quarkonium state, while in the CEM there is no constraint on the color state of the initial heavy-quark pair.
In NRQCD, in addition to the usual perturbative expansion in $\alpha_s$, a further expansion in $v$ (the relative velocity of the heavy-quark pair) is introduced.
As a consequence, one needs to take into account all the relevant Fock states of the heavy quark-antiquark pairs produced in the hard scattering. These Fock states are denoted by $^{2S+1}L_J^{(c)}$, where $S$ is  the spin of the pair, $L$ the orbital angular momentum, $J$ the total angular momentum and $c$ the color configuration, with $c=1, 8$. For an $S$-wave quarkonium state like the $J/\psi$ meson, the dominant contribution in the $v$ expansion, {\it i.e.}\ in the limit $v\to 0$, reduces to the traditional color singlet model~\cite{Berger:1980ni,Baier:1983va}. In addition, NRQCD predicts the existence of the color octet (CO) mechanism, according to which the $Q\overline Q$ pair can be produced at short distances also in CO states with different angular momentum and spin, and  subsequently evolves into the physical CS quarkonia by the nonperturbative emission of soft gluons. For an exhaustive and up-to-date overview on the phenomenology of quarkonium production see Ref.~\cite{Lansberg:2019adr} and references therein.

In this paper we focus on SSAs for $J/\psi$ production in $pp$ collisions within the GPM approach, studying in detail the role played by the CO w.r.t.~the CS mechanism within NRQCD, i.e.~we exploit the contributions from the $\leftidx{^{3}}{S}{_1}^{(1,8)}$,  $\leftidx{^{1}}{S}{_0}^{(8)}$ and $\leftidx{^{3}}{P}{_J}^{(8)}$ states. A complementary analysis within a TMD factorization scheme and still adopting the NRQCD framework in $ep\to J/\psi \, \rm{jet}\, X$ has been recently presented in Ref.~\cite{DAlesio:2019qpk}.

We consider all $2\rightarrow 1$ ($ab\to J/\psi$, CO mechanism) and $2\rightarrow 2$ ($ab\to J/\psi \,c$, CS and CO mechanisms) partonic subprocesses for $J/\psi$ production.
The amplitudes squared for these channels are calculated within NRQCD, and are found in agreement with the results in Refs.~\cite{Cho:1995vh,Cho:1995ce}. We will not report them here, referring the reader to these papers for their expressions.
We will show how the contribution from the quark Sivers function (potentially active in the CO terms) is still negligible. Concerning the access to the GSF, we find that some CO contributions turn out to be relevant, in particular when looking at backward rapidities. This could open a new way to access the gluon Sivers function.

We will then compare theoretical estimates for SSAs in $J/\psi$ production, obtained maximizing the gluon Sivers effect or adopting the GSF as extracted in Ref.~\cite{DAlesio:2018rnv}, against recent PHENIX data~\cite{Aidala:2018gmp}.
For completeness we will also consider in some detail the differential cross section within the GPM approach, comparing our results with PHENIX data~\cite{Adare:2009js}, in the moderate and low-$P_{T}$ (the transverse momentum of the $J/\psi$) region. As it will be shown, the inclusion of intrinsic transverse momentum effects could play an important role in this context.

As a last remark, we mention that the role of possible initial- and final-state interactions, and their interplay with the color octet mechanism, could open an interesting perspective in the study of SSAs within the CGI-GPM approach. This ongoing analysis will be presented in a future publication.

The paper is organized as follows: in Sec.~\ref{sec2} we present the calculation for $J/\psi$ production in $pp$ collisions within the GPM and NRQCD approaches. The corresponding numerical results, both for the unpolarized cross sections and the SSAs, are shown in Sec.~\ref{sec4}. Finally, in Sec.~\ref{sec5} we collect our conclusions.

%----------------------------------------------------------------------------------------
\section{Unpolarized cross section and single-spin asymmetry in $pp \to J/\psi\, X$: formalism}
\label{sec2}
We study the (un)polarized proton-proton collision process
\be
p^{\uparrow}(P_A)+p(P_B)\rightarrow J/\psi (P_h)+X \,,
\ee
where the arrow in the superscript indicates the transverse polarization of the proton with respect to the production plane, and the letters in round brackets represent the 4-momentum of the corresponding particle. We consider a center of mass frame where the two protons move along the $z$-axis, with the polarized proton along the positive $z$-axis, the $J/\psi$ is produced in the $x-z$ plane, and the $\uparrow$ transverse polarization is along $+y$. In the adopted NRQCD framework we have to consider the $2\rightarrow 1$ partonic subprocesses, namely $g+g\rightarrow J/\psi$ and $q+\bar{q}\rightarrow J/\psi$, and the $2\rightarrow 2$ ones, that is $g+g\rightarrow J/\psi+g$, $g+q(\bar q)\rightarrow J/\psi+q(\bar q)$ and $q+\bar{q}\rightarrow J/\psi+g$.

Within the GPM, the differential cross sections for proton-proton collisions are given, respectively for the $2\rightarrow 1$ and the $2\rightarrow 2$ channels, as
\begin{equation}\label{2d1}
 \begin{aligned}
E_h\frac{d^3\sigma^{2\to 1}}{d^3{\bm P}_h}={}&\frac{1}{2s}\sum_{a,b}\int \frac{dx_a}{x_a} \frac{dx_b}{x_b}  d^2{\bm k}_{\perp a}d^2{\bm k}_{\perp b} f_{a/p}(x_a, k_{\perp a})f_{b/p}(x_b, k_{\perp b})\frac{1}{(2\pi)^32}\\
& \times(2\pi)^4 \delta^4(p_a+p_b-P_h)|\mathcal{M}_{ab\rightarrow J/\psi }|^2\,,
\end{aligned}
\end{equation}
\begin{equation}\label{d221}
 \begin{aligned}
E_h\frac{d^3\sigma^{2\to 2}}{d^3{\bm P}_h}={}&\frac{1}{2s}\sum_{a,b,c}\int \frac{dx_a}{x_a} \frac{dx_b}{x_b}  d^2{\bm k}_{\perp a}d^2{\bm k}_{\perp b} f_{a/p}(x_a, k_{\perp a})f_{b/p}(x_b, k_{\perp b})\frac{1}{(2\pi)^32}\frac{d^3{\bm p}_c}{(2\pi)^32E_c}\\
& \times(2\pi)^4 \delta^4(p_a+p_b-P_h-p_c)|\mathcal{M}_{ab\rightarrow J/\psi c}|^2\,,
\end{aligned}
\end{equation}
where $a,b$ in the sum stand for $q=u,d,s$, $\bar q = \bar u, \bar d, \bar s$, and $g$.
After integrating over the phase space of the unobserved final parton $c$, Eq.~\eqref{d221} can be written
as
\begin{equation}\label{d2}
 \begin{aligned}
\frac{d\sigma^{2\to 2}}{dyd^2{\bm P_{T}}}=E_h\frac{d^3\sigma^{2\to 2}}{d^3{\bm P}_h}={}&
\frac{1}{2(2\pi)^2}\frac{1}{2s}\sum_{a,b,c}\int \frac{dx_a}{x_a} \frac{dx_b}{x_b}  d^2{\bm k}_{\perp
a}d^2{\bm k}_{\perp b} f_{a/p}(x_a,k_{\perp a})f_{b/p}(x_b,k_{\perp b})\delta(\hat{s}+\hat{t}+\hat{u}-M^2)\\
&\times|\mathcal{M}_{ab\rightarrow J/\psi c}|^2\,,
\end{aligned}
\end{equation}
where $M$ is the quarkonium mass and the $\mathcal{M}$'s are the hard scattering amplitudes, including the hadronization of the heavy-quark pair into the $J/\psi$. $\hat{s}$, $\hat{t}$ and $\hat{u}$ are the Mandelstam variables at the partonic level, while $P_{T}$ and $y$ are the transverse momentum and rapidity of the $J/\psi$ respectively.

For the $2\to 1$ channel, by exploiting the delta function and neglecting very small corrections of the order $k_\perp^2/s$, we can fix the light-cone momentum fractions $x_{a,b}$ as follows
\be
\delta^4 (p_a+p_b-P_h) = \frac{2}{s}\,\delta \bigg(x_a - \frac{E_h+P_L}{\sqrt{s}}\bigg)\, \delta \bigg(x_b - \frac{E_h-P_L}{\sqrt{s}}\bigg)\,
\delta^2 ({\bm k}_{\perp a} + {\bm k}_{\perp b} - \bm{P}_T)\,,
\ee
where $P_L$ is the $J/\psi$ longitudinal momentum. We can then rewrite Eq.~(\ref{2d1}) as
\begin{equation}
\label{2d1bis}
E_h\frac{d^3\sigma^{2\to 1}}{d^3{\bm P}_h}=\sum_{a,b}\frac{\pi}{x_a x_b s^2}\int d^2{\bm k}_{\perp a}d^2{\bm k}_{\perp b}
f_{a/p}(x_a, k_{\perp a})f_{b/p}(x_b, k_{\perp b})\,\delta^2 ({\bm k}_{\perp a} + {\bm k}_{\perp b} - \bm{P}_T) |\mathcal{M}_{ab\rightarrow J/\psi }|^2\,,
\end{equation}
where
\be
x_a = \frac{E_h+P_L}{\sqrt s} = \frac{M_T}{\sqrt{s}} e^y \hspace*{1cm}
x_b = \frac{E_h-P_L}{\sqrt s} = \frac{M_T}{\sqrt{s}} e^{-y},
\ee
with $M_T = \sqrt {P_T^2+M^2}$, the transverse mass of the $J/\psi$.

The SSA for the $p^\uparrow p\rightarrow J/\psi+X$ process is defined as usual as
\begin{equation}
\label{eq:AN}
A_N \equiv  \frac{\d \sigma^\uparrow - \d \sigma^\downarrow}{\d \sigma^\uparrow + \d \sigma^\downarrow}
 \equiv \frac{ \d\Delta\sigma}{ 2 \d\sigma}\,,
\end{equation}
where $\d \sigma^{\uparrow(\downarrow)}\equiv E_h \d^3\sigma^{\uparrow(\downarrow)}/\d^3\bm{P}_h$ indicates the transversely polarized cross section. The numerator of the asymmetry is sensitive to the Sivers function and the contributions from the $2\to 1$ and $2\to 2$ channels, which have to be added together (as for the corresponding unpolarized cross sections), read
\begin{equation}
\label{eq:A2d1}
\d\Delta\sigma^{2\to 1} = \sum_{a,b}\frac{\pi}{x_a x_b s^2}\int d^2{\bm k}_{\perp a}d^2{\bm k}_{\perp b} \,\Delta  \hat f_{a/{p^\uparrow}}(x_a, {\bm k}_{\perp a})\,f_{b/p}(x_b, k_{\perp b})\, \delta^2 ({\bm k}_{\perp a} + {\bm k}_{\perp b} - \bm{P}_T) |\mathcal{M}_{2\to 1}|^2,
\end{equation}
\be
\d\Delta\sigma^{2\to 2} = \frac{1}{2(2\pi)^2}\frac{1}{2s} \sum_{a,b,c}\int \frac{dx_a}{x_a} \frac{dx_b}{x_b}  d^2{\bm k}_{\perp a}d^2{\bm k}_{\perp b} \, \Delta  \hat f_{a/{p^\uparrow}}(x_a, {\bm k}_{\perp a})f_{b/p}(x_b, k_{\perp b})\,
\delta(\hat{s}+\hat{t}+\hat{u}-M^2)|\mathcal{M}_{2\to 2}|^2\,,
\ee
where $\Delta\hat f_{a/{p^\uparrow}}(x_a, {\bm k}_{\perp a})$ is the Sivers function of a parton with lightcone momentum fraction $x_a$ and transverse momentum $\bm{k}_{\perp a}={k}_{\perp a}(\cos\phi_a,\sin\phi_a,0)$, describing the azimuthal asymmetric distribution of an unpolarized parton in a transversely polarized proton with mass $M_p$.
Following the Trento conventions~\cite{Bacchetta:2004jz}, the Sivers function can be also expressed as
\begin{align}
\Delta \hat f_{a/\pup}\,(x_a, \bfk_{\perp a})  \,&\equiv \,
\hat f_{a/\pup}\,(x_a, \bfk_{\perp a}) - \hat f_{a/p^\downarrow}\,
(x_a, \bfk_{\perp a})\nonumber \\
\label{defsiv}
&= \,\Delta^N f_{a/\pup}\,(x_a, k_{\perp a}) \> \cos\phi_a\nonumber \\
&=  \,-2 \, \frac{k_{\perp a}}{M_p} \, f_{1T}^{\perp a} (x_a, k_{\perp a}) \>
\cos\phi_a \,.
\end{align}

\section{Numerical Results}
\label{sec4}
Following Refs.~\cite{DAlesio:2015fwo,DAlesio:2018rnv}, we adopt factorized Gaussian parameterizations for both the unpolarized TMD distribution and the Sivers function:
\be
f_{a/p}(x_a,k_{\perp a}) = \frac{e^{-k_{\perp a}^2/\langle k_{\perp a}^2 \rangle}}{\pi \langle k_{\perp a}^2\rangle} f_{a/p}(x_a)
\ee
and
\begin{equation}
\Delta^N\! f_{a/p^\uparrow}(x_a,k_{\perp a}) =   \left (-2\frac{k_{\perp a}}{M_p}  \right )f_{1T}^{\perp\,a}
(x_a,k_{\perp a})  = 2 \, {\cal N}_a(x_a)\,f_{a/p}(x_a)\,
h(k_{\perp a})\,\frac{e^{-k_{\perp a}^2/\langle k_{\perp a}^2 \rangle}}
{\pi \langle k_{\perp a}^2 \rangle}\,,
\label{eq:siv-par-1}
\end{equation}
where $f_{a/p}(x_a)$ is the usual collinear parton distribution and
\begin{equation}
{\cal N}_a(x_a) = N_a x_a^{\alpha}(1-x_a)^{\beta}\,
\frac{(\alpha+\beta)^{(\alpha+\beta)}}
{\alpha^{\alpha}\beta^{\beta}}\,,
\label{eq:nq-coll}
\end{equation}
with $|N_a|\leq 1$ and
\begin{equation}
h(k_{\perp a}) = \sqrt{2e}\,\frac{k_{\perp a}}{M'}\,e^{-k_{\perp a}^2/M'^2}\,.
\label{eq:h-siv}
\end{equation}
This ensures that the Sivers function satisfies the positivity bound for all values of $x_a$ and $k_{\perp a}$:
\begin{equation}
\vert \Delta^N f_{a/\pup}\,(x_a, k_{\perp a}) \vert  \le 2\,f_{a/p}\,(x_a, k_{\perp a})
\,,~~{\mathrm{ or}}~~
\frac{k_\perp}{M_p}\, \vert f_{1T}^{\perp a} (x_a, k_{\perp a})\vert \le  f_{a/p}\,(x_a, k_{\perp a})~.
\end{equation}
If we define the parameter
\begin{equation}
\rho_a = \frac{M'^2}{\langle k_{\perp a}^2 \rangle +M'^2}\, ,
\label{eq:rho}
\end{equation}
such that $0< \rho_a < 1$, then Eq.~(\ref{eq:siv-par-1}) becomes
\begin{equation}
\label{eq:siv-par}
\Delta^N\! f_{a/p^\uparrow}(x_a,k_{\perp a}) =   \frac{\sqrt{2e}}{\pi}   \,2\, {\cal N}_a(x_a)\, f_{a/p}(x_a) \, \sqrt{\frac{1-\rho_a}{\rho_a}}\,k_{\perp a}\, \frac{e^{-k_{\perp a}^2/ \rho_a\langle k_{\perp a}^2 \rangle}}
{\langle k_{\perp a}^2 \rangle^{3/2}}~.
\end{equation}

For the collinear unpolarized parton distributions we will adopt the CTEQL1 set~\cite{Pumplin:2002vw}, at the factorization scale equal to $M_T$, adopting the DGLAP evolution equations.

By using the above factorized parameterizations we could integrate analytically the expressions entering the numerator and the denominator of $A_N$ for the $2\to 1$ channels, as follows:
\be
2d\sigma^{2\to 1} = \frac{1}{s^2}\sum_{a,b}\frac{1}{x_a x_b} \frac{1}{\langle k_{\perp a}^2 \rangle + \langle k_{\perp b}^2 \rangle} \exp\!\bigg(\!-\frac{P_T^2}{\langle k_{\perp a}^2 \rangle + \langle k_{\perp b}^2 \rangle}\bigg)  2\,f_{a/{p}}(x_a)\,f_{b/p}(x_b)\, |\mathcal{M}_{2\to 1}|^2,
\ee
\be
 d\Delta\sigma^{2\to 1} = \frac{\sqrt{2e}}{s^2}\sum_{a,b}\frac{1}{x_a x_b}\,
 \frac{\sqrt{\rho_a^3 (1-\rho_a)\langle k_{\perp a}^2\rangle}}{(\rho_a\langle k_{\perp a}^2 \rangle + \langle k_{\perp b}^2 \rangle)^2} P_T \exp\!\bigg(\!-\frac{P_T^2}{\rho_a\langle k_{\perp a}^2 \rangle + \langle k_{\perp b}^2 \rangle}\bigg) 2 \, {\cal N}_a(x_a)\, f_{a/{p}}(x_a)\,f_{b/p}(x_b)\, |\mathcal{M}_{2\to 1}|^2\,.
\label{eq:2to1}
\ee
The other cases for the $2\to 2$ channel will have to be integrated numerically.

\subsection{Unpolarized cross sections for $J/\psi$ production}
\label{unpxs}

In order to give theoretical estimates for the unpolarized cross sections we have to fix all parameters entering our expressions, namely the Gaussian widths for the TMDs (both for quarks and gluons) and the LDMEs for the hadronization of the heavy-quark pair into a $J/\psi$.

For the unpolarized quark TMDs we adopt the Gaussian width as extracted in Ref.~\cite{Anselmino:2005nn}, that is $\langle k_{\perp q}^2 \rangle= 0.25$ GeV$^2$, while for the gluon TMD a reasonably good description of unpolarized cross section data can be obtained with the value adopted in Ref.~\cite{DAlesio:2017rzj}, that is $\langle k_{\perp g}^2 \rangle= 1$ GeV$^2$ (see below). %\textbf{We have checked the other reasonable values around this give very similar results.}

Concerning the LDME choice, some comments are in order: these quantities have to be extracted from fits to data, and despite their expected universality, different sets are available in the literature. The reason basically traces back to the theoretical assumptions adopted in their extractions (like the accuracy of the pQCD calculation), the selected data set and the imposed kinematical cuts, mainly on $P_T$. In particular, most of the analyses carried out within NRQCD consider only moderate to large $P_T$ data, namely $P_T>3-5$ GeV (or even larger). This is somehow necessary in a fixed order calculation in pQCD, since in a collinear factorization scheme, the differential cross section at $P_T=0$ manifests infrared divergences, when the production of CO states is included. This indeed happens for the states $\leftidx{^{1}}{S}{_0}^{(8)}$ and $\leftidx{^{3}}{P}{_J}^{(8)}$ because of their Feynman diagram topology.

Some attempts to describe the low-$P_T$ region have been made, like in the color glass condensate approach~\cite{Kang:2013hta,Ma:2014mri}, which is able to match the behaviour of pure collinear NRQCD in the region where the two approaches overlap (namely, around 2-3 GeV). Further analyses have been performed adopting the $k_T$-factorization formalism in NRQCD~\cite{Baranov:2002cf} or in the color evaporation model~\cite{Cheung:2018tvq}, or employing, still within NRQCD, the Collins-Soper-Sterman (CSS) approach~\cite{Berger:2004cc,Sun:2012vc}, that allows to resum soft gluon contributions.

In our TMD approach within NRQCD, we do not include any higher-order correction or any kind of resummation.
Intrinsic transverse momentum effects, then, act as an effective way to \emph{cure} the divergences in the unpolarized cross sections at low $P_T$. As a result, the differential cross section is finite at $P_T=0$ and, as shown below, in reasonable agreement with data, within the same picture adopted for the SSAs (our main goal).

We have then checked several LDME sets, looking in particular at those extracted in a region sufficiently close to (or overlapping with) the one we are interested in, that is at small to moderate $P_T$ values.
Among them, we have selected the following two sets.
One (named BK11 in the following), extracted in collinear NRQCD at NLO~\cite{Butenschoen:2011yh}, is based on a global fit including $pp$ data down to $P_T$ around 2-3 GeV, for both direct and prompt $J/\psi$ production. %\textbf{The last one set subtracting or not subtracting the so-called feed-down contribution}.
%\footnote{\textbf{We note that they compute only direct production, as we are doing, without subtracting the so-called feed-down contribution, that in principle enters in a subset of data.}}.
We will use the default set, obtained fitting hadroproduction data for $P_T>3$ GeV. We have checked that other related sets~\cite{Butenschoen:2012qh}, with different $P_T$ cuts, or obtained subtracting the feed-down contributions, give very similar results.

The second one (referred to as SYY13) was obtained employing the CSS approach within NRQCD~\cite{Sun:2012vc} and considering $pp$ data in the very low-$P_T$ region. In this analysis, in order to control the infrared behaviour of the perturbative part in the impact parameter space, the authors adopt the so-called $b_{\rm max}$ prescription, with $b_{\rm max} = 0.5$ GeV$^{-1}$. Another aspect of this extraction is that since the color singlet contribution is not affected by divergences in the low-$P_T$ region, it does not need any soft gluon resummation. For this reason it was not included at all in the fit\footnote{We thank Feng Yuan for clarifying this point.}. In this respect, this set has to be consistently used only for the CO channels. We will come back to these points when discussing our results below.
It is also worth noticing that our TMD approach is somehow an effective way of resumming soft gluons to regulate the most singular terms in the partonic subprocesses. We could therefore have a direct comparison in the same kinematical region and within a somehow related approach.
Last but not least, we note that, while different LDME sets can give very different values for the unpolarized cross sections, the SSAs, being ratios of cross sections, are much less sensitive to them.

\begin{figure}[th]
\begin{center}
\includegraphics[trim = 1.cm 0cm 1cm 0cm, width=8.5cm]{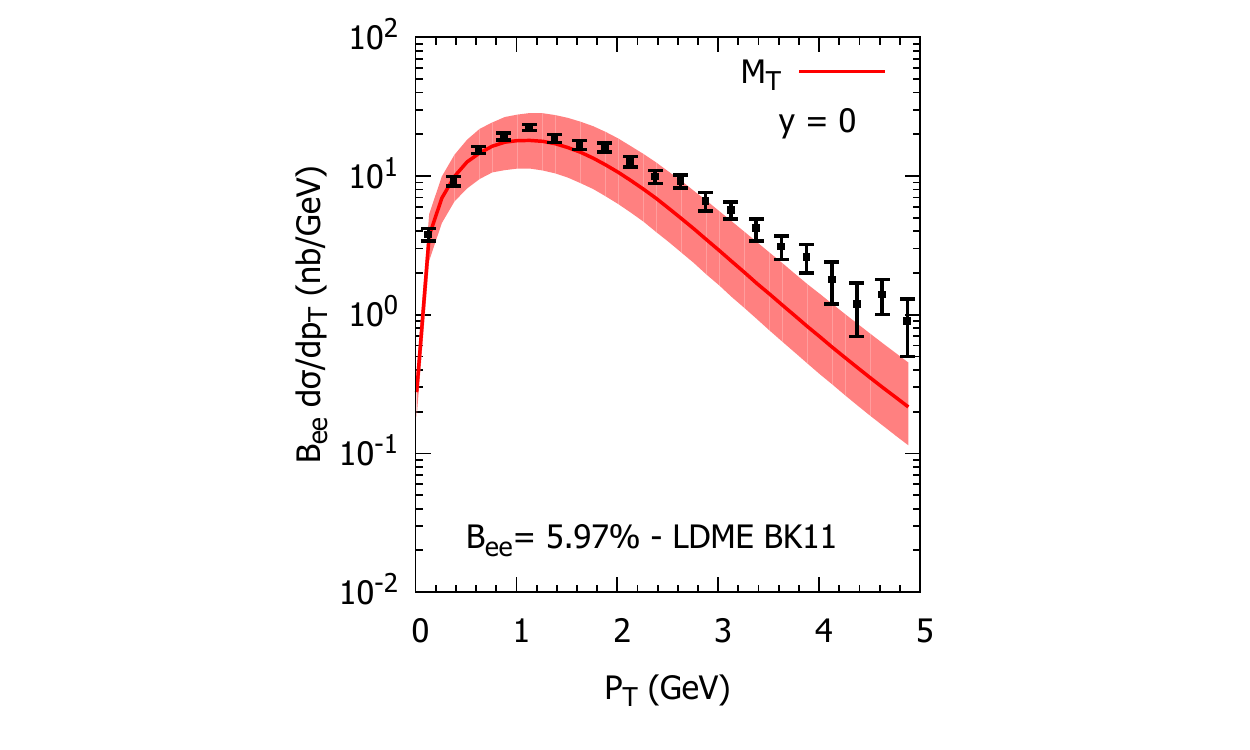}
\includegraphics[trim = 1.cm 0cm 1cm 0cm, width=8.5cm]{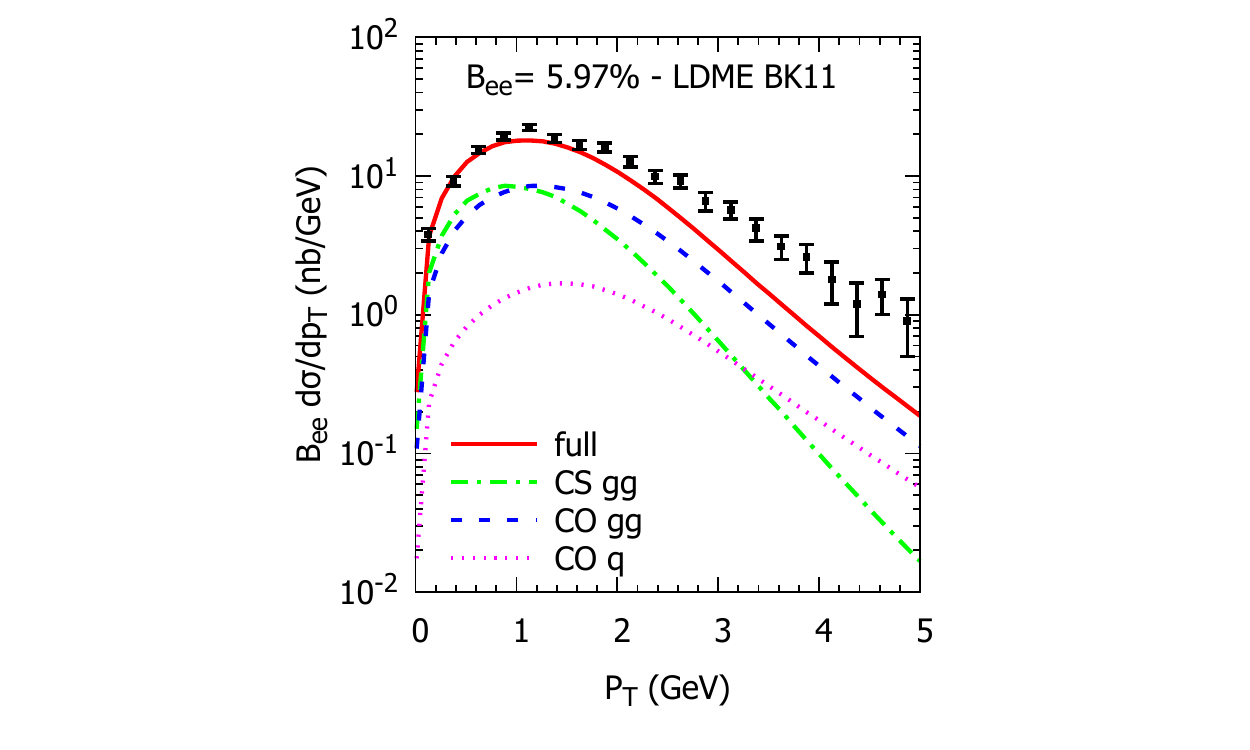}
\caption{Left panel: Unpolarized cross section for the process $p p\to J/\psi\, X$, as a function of $P_T$ at $\sqrt s=200$ GeV and $y=0$, within the GPM approach compared with PHENIX data~\cite{Adare:2009js}, obtained with the BK11 LDME set~\cite{Butenschoen:2011yh}. The band represents the theoretical uncertainty due to variation of the factorization scale from $M_T/2$ to $2M_T$.
Right panel: separate contributions to the full cross section (red solid line) for $\mu=M_T$ from the color singlet state (green dot-dashed line) and the color octet states for the gluon-gluon (blue dashed line), gluon-quark, quark-gluon and quark-antiquark (magenta dotted line) initiated subprocesses.}
\label{fig1:unpol}
\end{center}
\end{figure}

In Fig.~\ref{fig1:unpol} (left panel) we show our estimates, obtained with the BK11 LDME set, of the unpolarized cross section for $J/\psi$ production in $pp$ collisions, compared against PHENIX data~\cite{Adare:2009js}, at $\sqrt s = 200$ GeV and rapidity $y=0$, as a function of $P_T$. The band represents the theoretical uncertainty due to variation of the factorization scale $\mu$ from $M_T/2$ to $2M_T$. The inclusion of transverse momentum effects is clearly able to remove the singularities coming from the CO contributions  in the small-$P_T$ region and the agreement with data is good enough for our purposes. For completeness, we have to mention that a residual small instability is still present for $P_T\le 1$-2 GeV, where one could probe very small values of the partonic Mandelstam invariants ($\hat t, \hat u$) in the hard partonic cross sections. To control it, we have introduced an infrared regulator, $\mu_{\rm IR}$, cutting out final partonic transverse momenta (in the partonic center of mass frame) lower than $\mu_{\rm IR}$. We have checked that a regulator around 0.8-1 GeV is enough. Our results are shown for $\mu_{\rm IR}=0.8$ GeV.

\begin{figure}[h!]
\begin{center}
\includegraphics[trim = 1.cm 0cm 1cm 0cm, width=8.5cm]{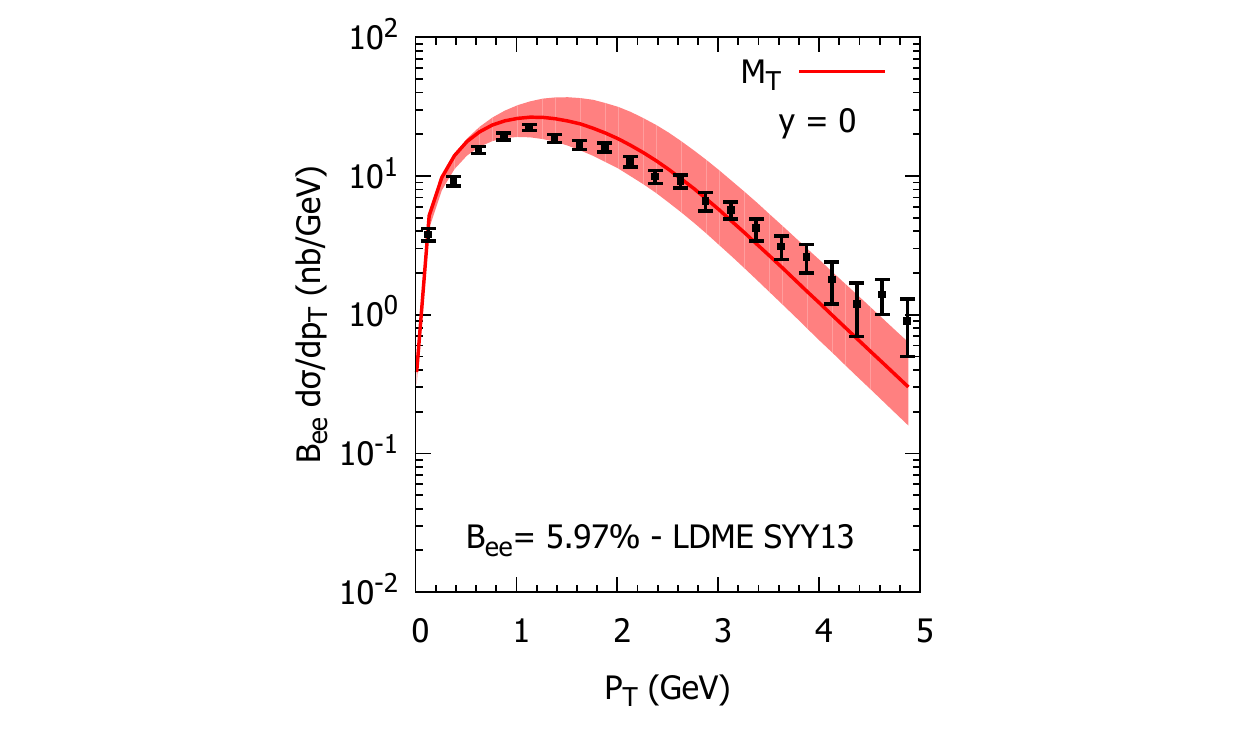}
\includegraphics[trim = 1.cm 0cm 1cm 0cm, width=8.5cm]{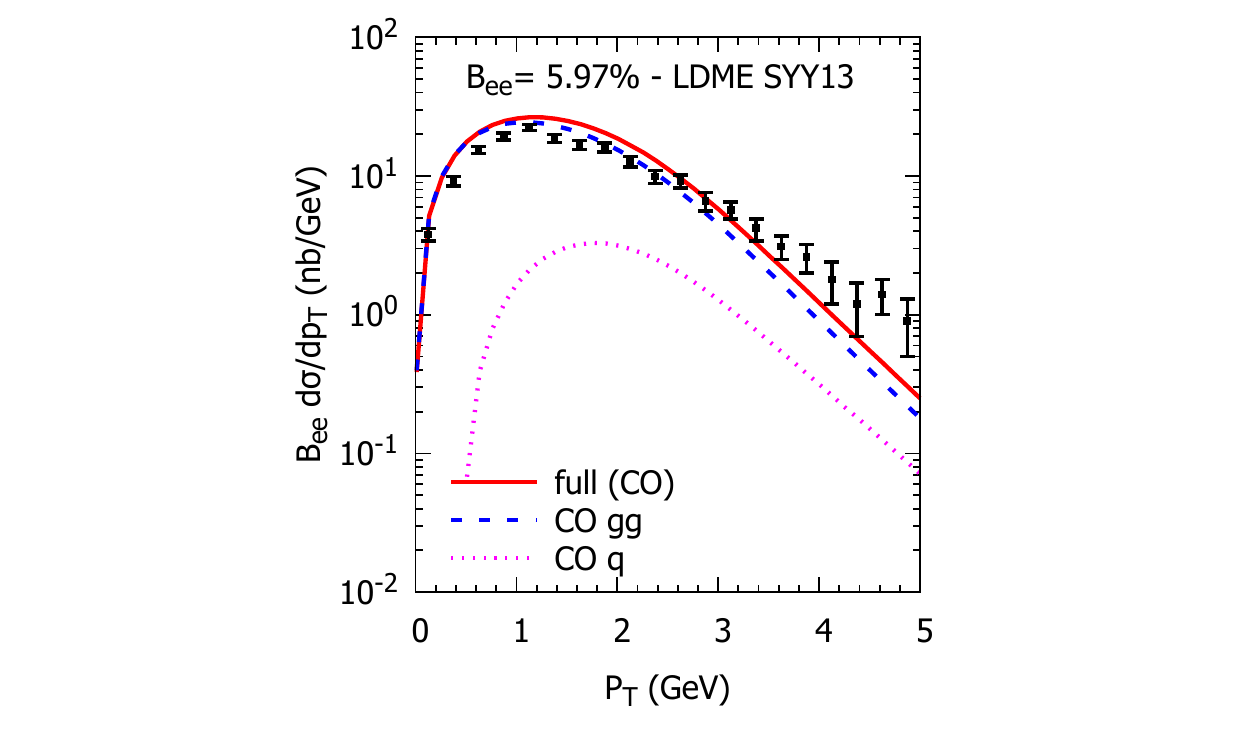}
\caption{Left panel: Unpolarized cross section for the process $p p\to J/\psi\, X$ as a function of $P_T$ at $\sqrt s=200$ GeV and $y=0$, within the GPM approach compared with PHENIX data~\cite{Adare:2009js}, obtained with the SYY13 LDME set~\cite{Sun:2012vc}. The band represents the theoretical uncertainty due to variation of the factorization scale from $M_T/2$ to $2M_T$.
Right panel: separate contributions to the full cross section (red solid line) for $\mu=M_T$ from the color octet states for the gluon-gluon (blue dashed line), gluon-quark, quark-gluon and quark-antiquark (magenta dotted line) initiated subprocesses.}
\label{fig2:unpol}
\end{center}
\end{figure}

In the right panel of Fig.~\ref{fig1:unpol} we show the different contributions to the cross section at $\mu=M_T$. As one can see, in the small-$P_T$ region the CS (blue dashed line) and the CO (green dot-dashed line) gluon-gluon fusion NRQCD terms are comparable, while at larger $P_T$ values the CO mechanism turns to be dominating. More important, from our point of view, is the almost negligible role played by CO subprocesses involving a quark in the initial state (magenta dotted line). On this basis, we can still say that in this region $J/\psi$ production is dominated by gluon initiated subprocesses and is then an ideal tool to access the gluon Sivers function.
It is also worth to remark that for the CO mechanism the dominant channels are the $2\to 2$ ones, while the $2\to 1$ $gg\to J/\psi$ channel plays some role only in the low-$P_T$ region and represents, at most, 20\% of the full $gg$ contribution (the $q\bar q$ channel being totally negligible).

In Fig.~\ref{fig2:unpol} (left panel) we show the corresponding estimates, obtained with the SYY13 LDME set, of the unpolarized cross section for $J/\psi$ production in $pp$ collisions. Notice that for this set, for consistency, no CS contribution has been included. We have nevertheless checked that, at variance with the BK11 set, it would be completely negligible.
In this case, in order to control the IR instabilities, and get closer to data, we choose a larger value for $\mu_{\rm IR}$.
This is somehow related to the fact that this LDME set has been extracted from a fit to very low-$P_T$ data adopting, as an IR regulator, a relatively small $b_{\rm max}$ value. In our scheme, by choosing values around 1-1.3 GeV we get a reasonable description of data. Results are shown for $\mu_{\rm IR}= 1.2$ GeV.

In the right panel of Fig.~\ref{fig2:unpol} we show the different contributions to the cross section at $\mu=M_T$.
Again the subprocesses involving a quark in the initial state (magenta dotted line) are almost negligible. Notice that in this case the $2\to 1$ channel, $gg\to J/\psi$, is even bigger than the $2\to 2$ one in the small-$P_T$ region (60-70\% of the full $gg$ contribution), being around 20\% at $P_T\sim 2$ GeV and becoming negligible at higher $P_T$.

%------------------------------
\subsection{Single-spin asymmetry in $p^\uparrow p\to J/\psi\, X$}
\label{ssas}

We now move to the main issue of our study: the phenomenological analysis of SSAs for $J/\psi$ production.
Concerning the Sivers effect, we will start considering its maximized contribution to $A_N$, in order to exploit the potential differences coming from the production mechanism. More precisely, we will adopt $\rho_{q,g}=2/3$ and ${\cal N}_{q,g}(x)=\pm 1$ in Eq.~(\ref{eq:siv-par}). This indeed allows to test directly the role of the Sivers azimuthal phase through the dynamics and kinematics at the parton level.

\begin{figure}[b!]
\begin{center}
\includegraphics[trim = 1.cm 0cm 1cm 0cm, width=8.5cm]{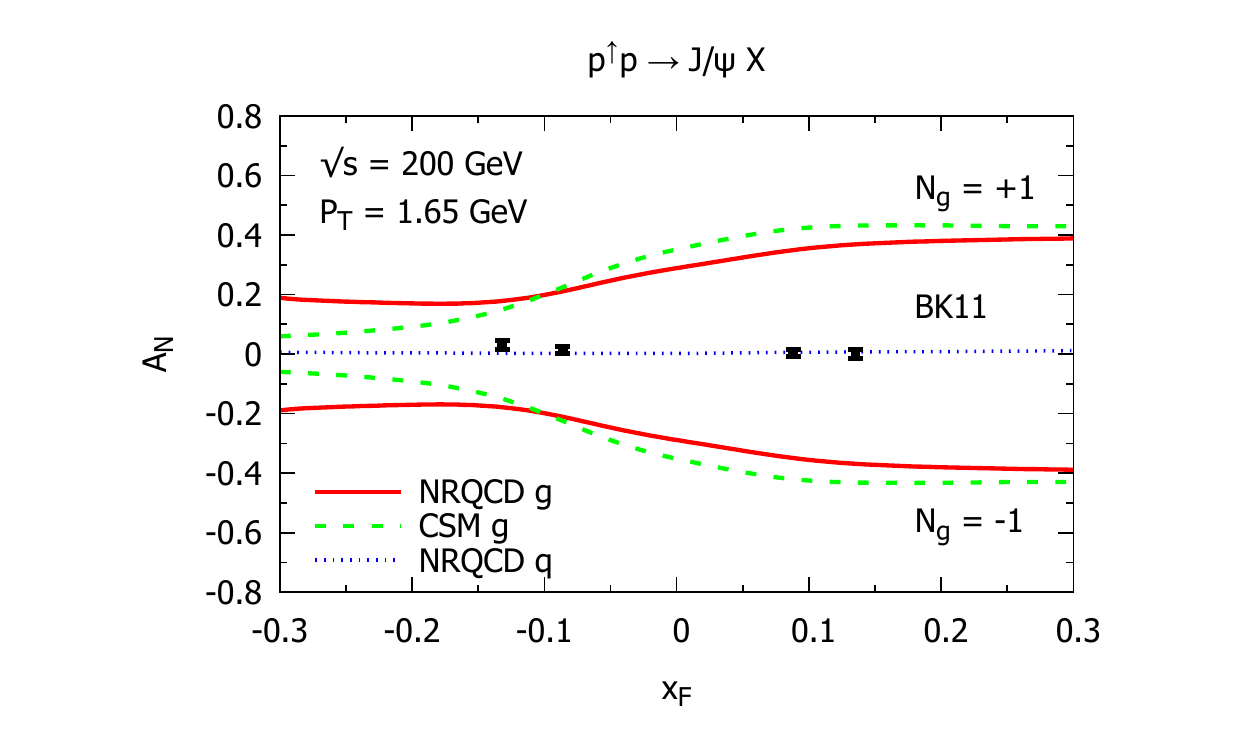}
\includegraphics[trim = 1.cm 0cm 1cm 0cm, width=8.5cm]{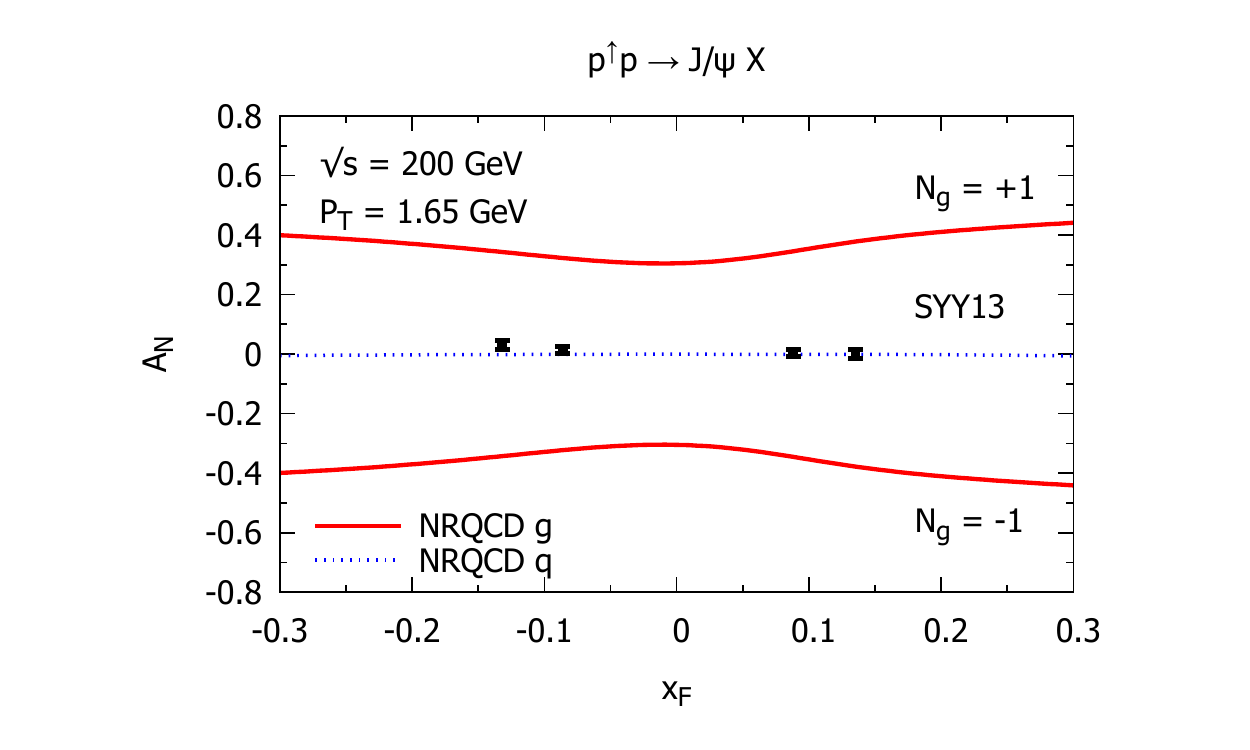}
\caption{Maximized values for $A_N$ for the process $p^\uparrow p\to J/\psi\, X$  as a function of $x_F$ at $\sqrt s=200$ GeV and $P_T=1.65$ GeV, for the gluon Sivers effect both in the CSM (green dashed lines) and in NRQCD (red solid line), and for the quark Sivers effect (blue dotted line), adopting the BK11 (left panel) and the SYY13 (right panel) LDME sets. Notice that for the SYY13 set only CO states are included (see text for further details). Data are taken from \cite{Aidala:2018gmp}.}
\label{fig:ANsatxF}
\end{center}
\end{figure}

In Fig.~\ref{fig:ANsatxF}, we show the maximized contributions to $A_N$, coming separately from the gluon Sivers effect, in the CSM (green dashed line) and in NRQCD (red solid line), and from the quark Sivers effect, only in NRQCD, (blue dotted line), adopting the BK11 (left panel) and the SYY13 (right panel) LDME sets.
Notice that the CSM results, which do not depend on the LDME set, supersede the corresponding ones in Ref.~\cite{DAlesio:2018rnv} (Fig.~6, left panel), affected by a small error in the numerical code.

As one can see, only the gluon contribution could be sizeable. We also notice that in the forward region the CSM and the NRQCD estimates are very similar, while in the backward region the CO terms could, in principle, give a potentially large contribution, very important for the SYY13 set. This issue deserves some comments (see also the discussion of the relative role of $2\to 1$ vs $2\to 2$ channels in Sec.~\ref{unpxs}).
The $2\to 2$ channels are suppressed in the backward region due to the integration over the Sivers azimuthal phase, which for $x_F<0$ plays a less effective role in the hard parts. On the contrary, the $2\to 1$ channels, only CO terms (see Eq.~\ref{eq:2to1}), for which we have integrated over the partonic variables, are not affected by this. Their hard parts do not depend on the $\hat t$ and $\hat u$ Mandelstam invariants and the Sivers azimuthal phase enters, besides its explicit $\cos\phi$ dependence, only in the delta function on transverse momenta (see Eq.~(\ref{eq:A2d1})). This implies that there is no correlation between this phase dependence and $x_F$ (through the Mandelstam invariants) and therefore the $2\to 1$ contributions are symmetric in $x_F$, as happens for the unpolarized case.
Moreover, for the SYY13 set, the gluon Sivers effect in the backward region turns out to be comparable to the one in the forward region. This opens a new way to access the GSF in quarkonium production, since also the backward region could be useful to put a constraint on this important TMD. From the data it is clear that only a GSF much smaller than its positivity bound could lead to reasonable results.

We now consider some results for the dominant effect, adopting for the GSF the parametrization extracted in Ref.~\cite{DAlesio:2018rnv}, where the best fit parameters are
\begin{equation}
\label{parGSF}
N_g=0.25, ~~~~~~~\alpha=0.6,~~~~~~~ \beta=0.6, ~~~~~~~\rho=0.1, ~~~~~~~[\langle k_{\perp g}^2\rangle = 1 \;{\rm GeV}^2]\,.
\end{equation}

%--------------------------------------
\begin{figure}[t]
\begin{center}
\includegraphics[trim = 1.cm 0cm 1cm 0cm, width=8.5cm]{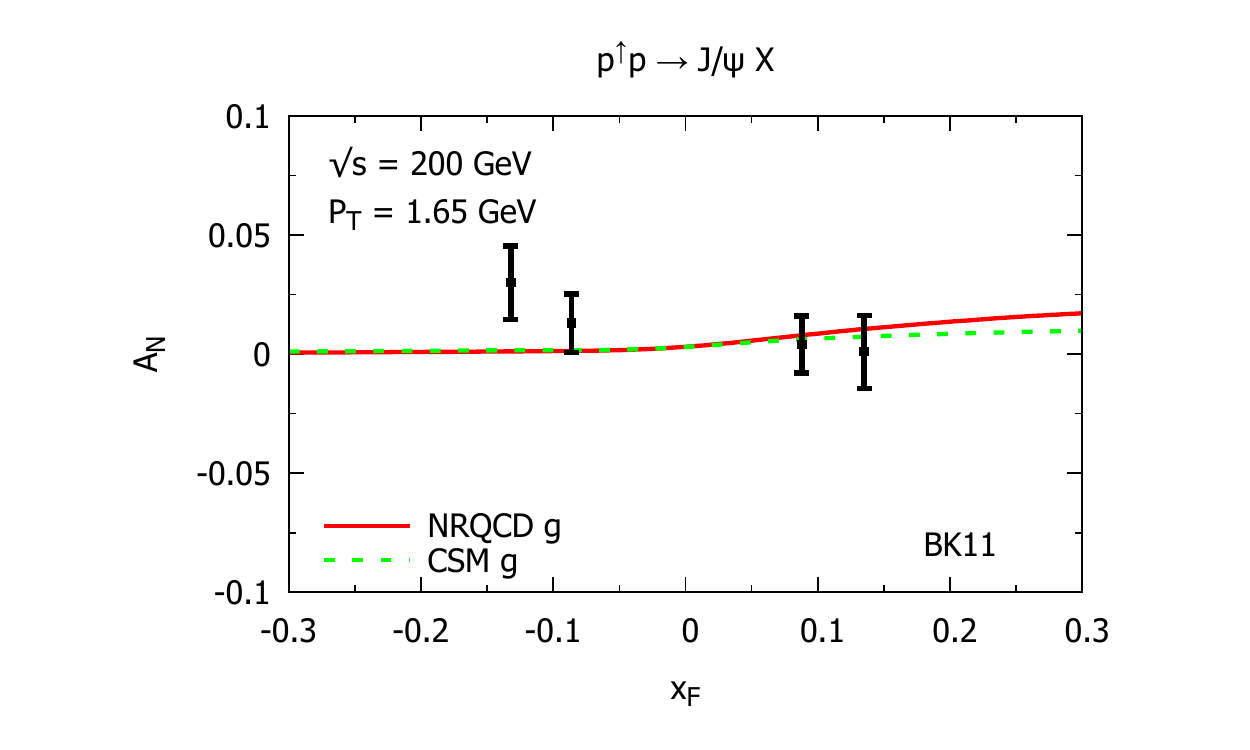}
\includegraphics[trim = 1.cm 0cm 1cm 0cm, width=8.5cm]{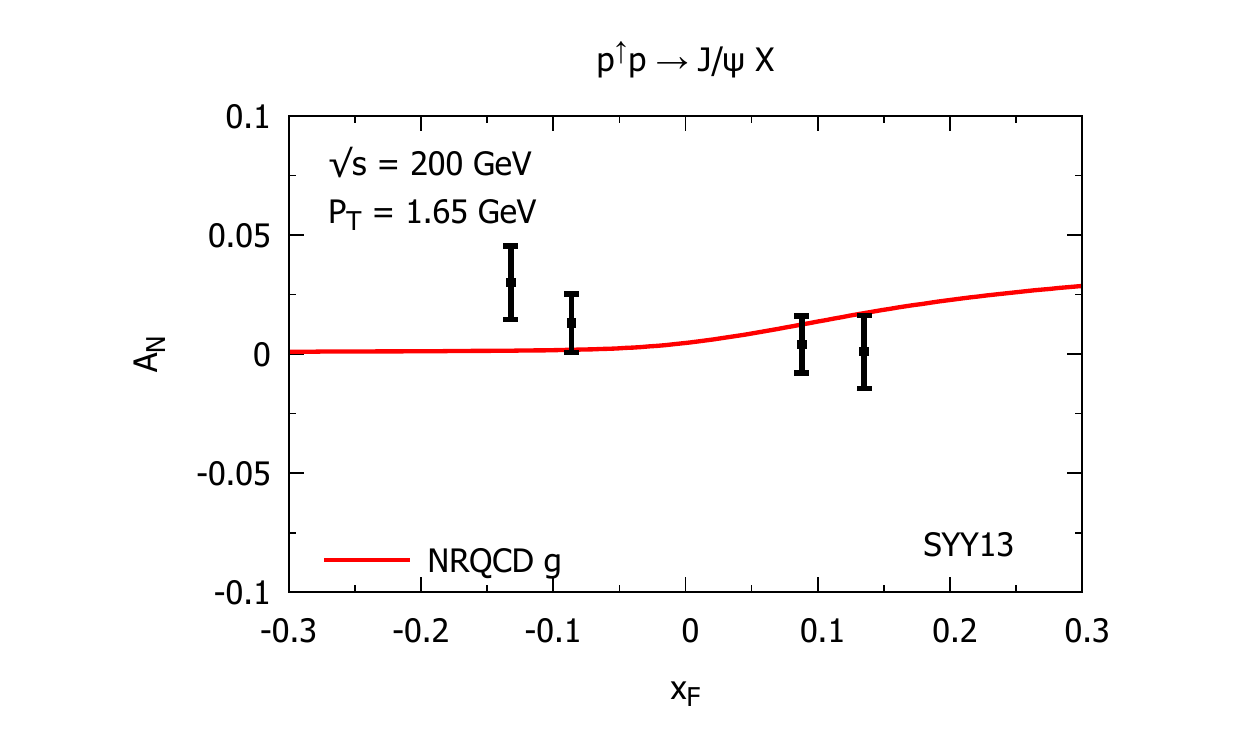}
\caption{Estimates of $A_N$ for the process $p^\uparrow p\to J/\psi\, X$ as a function of $x_F$  at $\sqrt s=200$ GeV and $P_T=1.65$ GeV, both in the CSM (green dashed line) and the NRQCD approach (red solid line), adopting the BK11 (left panel) and the SYY13 LDME set (right panel), compared against PHENIX data~\cite{Aidala:2018gmp}. The curves are calculated adopting the parameters in Eq.~(\ref{parGSF}) for the GSF. }
\label{fig:ANxF}
\end{center}
\end{figure}

\begin{figure}[b]
\begin{center}
\includegraphics[trim = 1.cm 0cm 1cm 0cm, width=8.5cm]{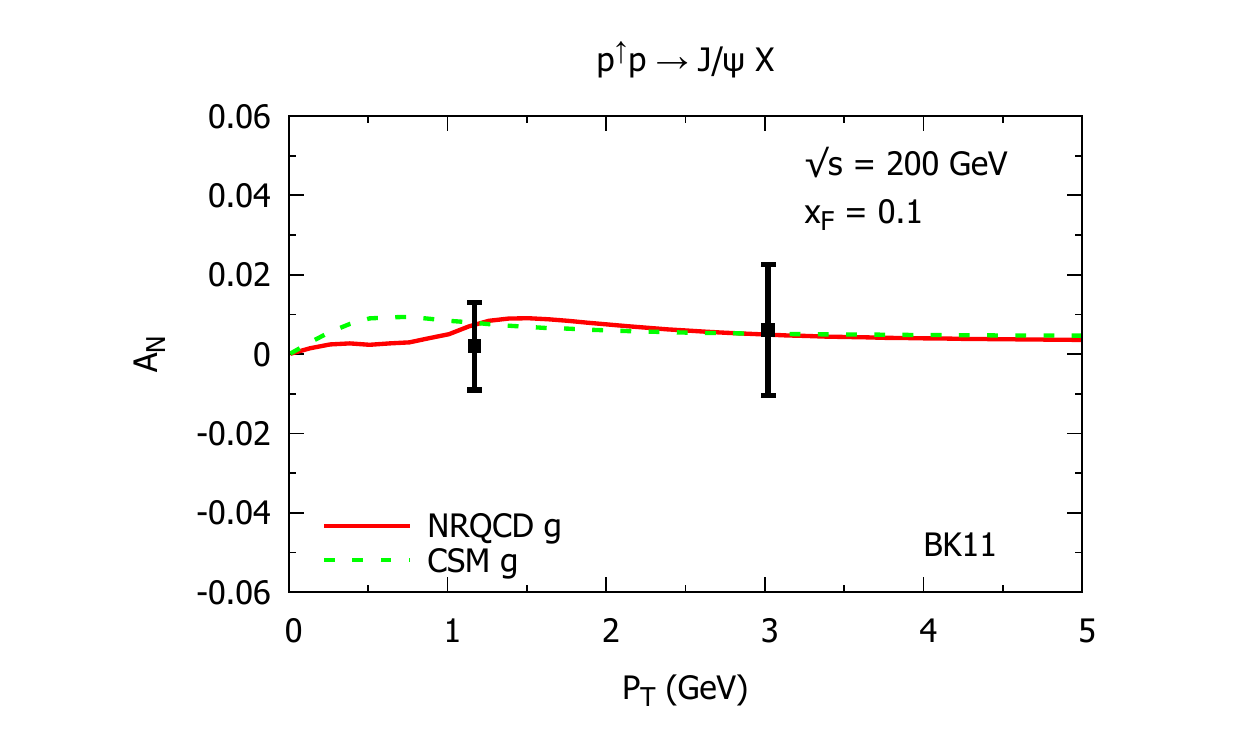}
\includegraphics[trim = 1.cm 0cm 1cm 0cm, width=8.5cm]{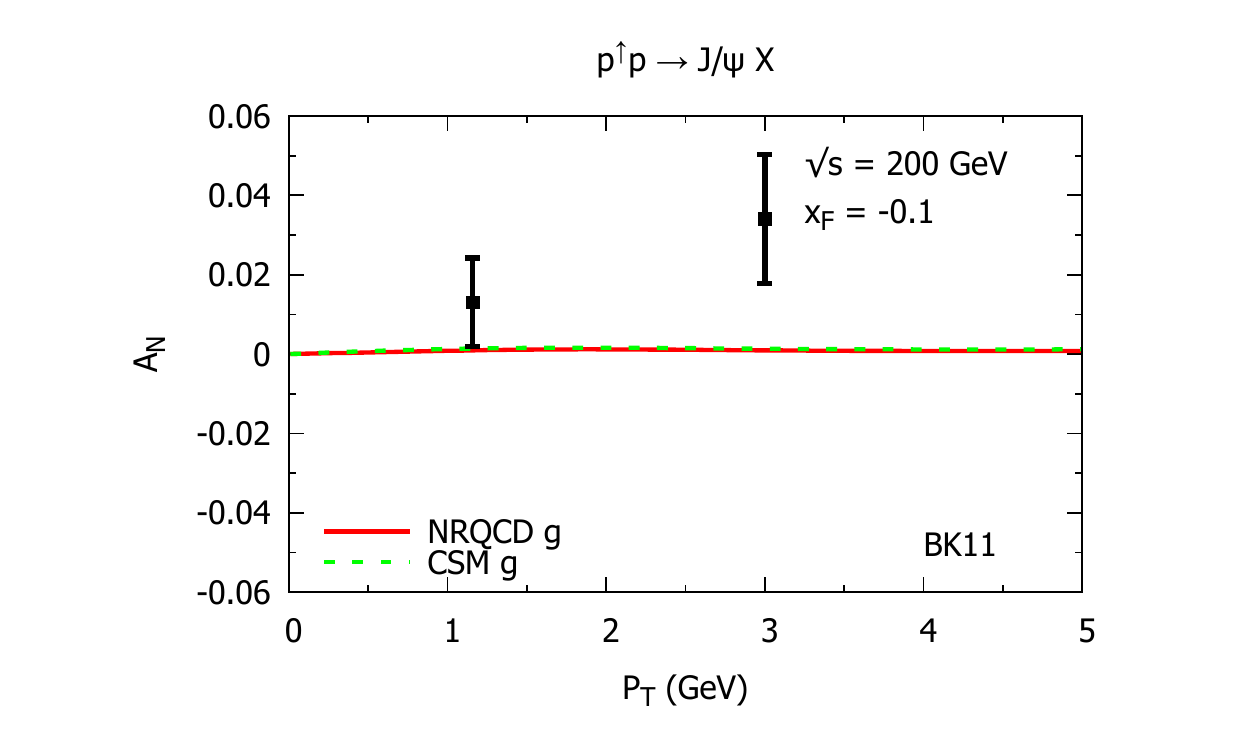}
\caption{Estimates of $A_N$ for the process $p^\uparrow p\to J/\psi\, X$ as a function of $P_T$ at $\sqrt s=200$ GeV and $x_F=0.1$ (left panel) and $x_F=-0.1$ (right panel), both in the CSM (green dashed line) and the NRQCD approach (red solid line), adopting the BK11 set, compared against PHENIX data~\cite{Aidala:2018gmp}. The curves are calculated adopting the parameters in Eq.~(\ref{parGSF}) for the GSF. }
\label{fig:ANpTBK11}
\end{center}
\end{figure}

In Fig.~\ref{fig:ANxF} we show a comparison with PHENIX data~\cite{Aidala:2018gmp} of our estimates for $A_N$ for $J/\psi$ production as a function of $x_F$, at $\sqrt s=200$ GeV and $P_T=1.65$ GeV, both in the CSM (green dashed line) and the NRQCD approach (red solid line), adopting the BK11 (left panel) and the SYY13 (right panel) LDME sets. The curves are obtained adopting the parameters in Eq.~(\ref{parGSF}) for the GSF. In Fig.~\ref{fig:ANpTBK11} we present the corresponding estimates for $A_N$ as a function of $P_T$ at $\sqrt s=200$ GeV and $x_F=0.1$ (left panel) and $x_F=-0.1$ (right panel), both in the CSM (green dashed line) and the NRQCD approach (red solid line), adopting the BK11 set, compared against PHENIX data~\cite{Aidala:2018gmp}. Similar results, not shown, are obtained adopting the SYY13 set. In such cases the sensitivity to the quarkonium production mechanism is negligible, being the predictions almost indistinguishable. Moreover, the discrepancy in the description of the most backward data point is present both in the CSM and in NRQCD.

%--------------------------------------

For its potential role in this context, we consider the corresponding $A_N$ in $J/\psi$ production for the kinematics reachable at LHC in the fixed target mode with a transversely polarized target (see the AFTER~\cite{Brodsky:2012vg,Hadjidakis:2018ifr} and LHCSpin~\cite{DiNezza:2019ziv,Aidala:2019pit} proposals at CERN). In such a configuration one could probe even larger light-cone momentum fractions in the polarized proton, accessing the gluon TMDs in a very interesting and complementary region.

\begin{figure}[t]
\begin{center}
\includegraphics[trim = 1.cm 0cm 1cm 0cm, width=8.5cm]{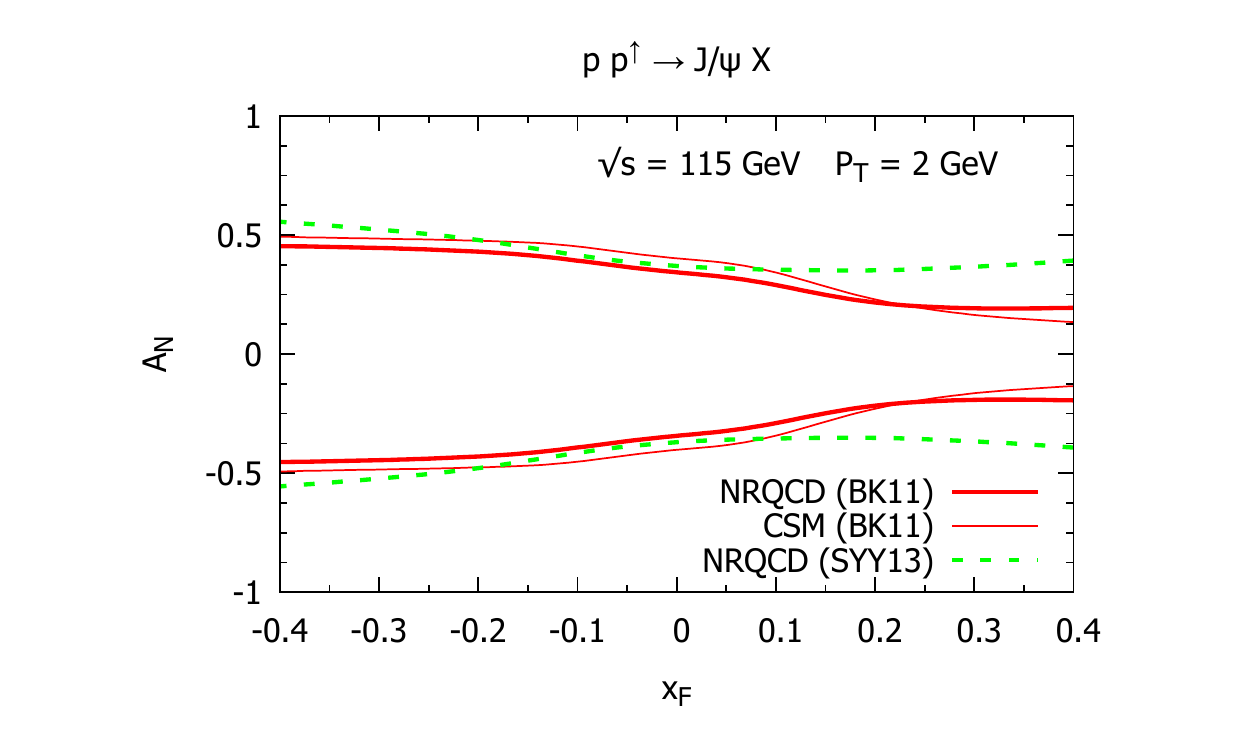}
\includegraphics[trim = 1.cm 0cm 1cm 0cm, width=8.5cm]{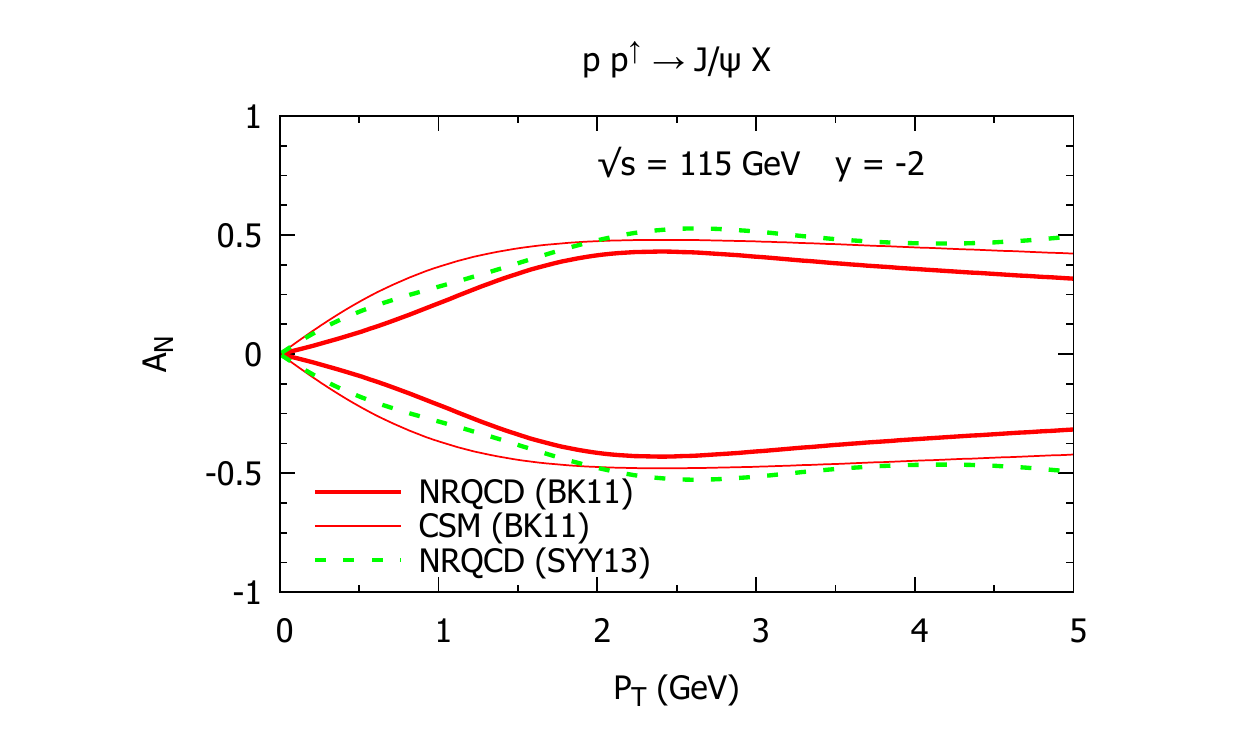}
\caption{Maximized values for $A_N$ for the process $p p^\uparrow \to J/\psi\, X$ at $\sqrt s=115$ GeV and $P_T=2$ GeV as a function of $x_F$ (left panel) and at $y=-2$ as a function of $P_T$ (right panel), for the gluon Sivers effect only. 
Curves are obtained within the NRQCD approach, adopting the BK11 (thick red solid lines) and the SYY13 (green dashed lines) LDME sets and in the CSM (thin red solid lines). Notice that here negative rapidities correspond to the forward region for the polarized proton.}
\label{fig:ANsat_LHCb}
\end{center}
\end{figure}

In Fig.~\ref{fig:ANsat_LHCb} we present our maximized estimates for $A_N$ for $p p^\uparrow \to J/\psi\, X$ at $\sqrt s=115$ GeV, at fixed $P_T=2$~GeV, as a function of $x_F$ (left panel) and at fixed rapidity $y=-2$, as a function of $P_T$ (right panel). Notice that in such a configuration the backward rapidity region refers to the forward region for the polarized proton target. In particular, we show our results both adopting the BK11 set, NRQCD (thick red solid lines) and CSM (thin red solid lines) contributions, and the SYY13 set, only CO term (green dashed lines). No significant differences appear also in this case. Notice that the predictions obtained adopting the GSF parametrization of Eq.~(\ref{parGSF}) would give results compatible with zero.

Also in this case, the CSM results, independent of the adopted LDME set, supersede the corresponding ones in Ref.~\cite{DAlesio:2018rnv} (Fig.~8).

\section{Conclusions}\label{sec5}
In this paper, extending previous work within the color singlet model, we have considered $J/\psi$ production in proton-proton collisions within a TMD scheme, which allows to include spin and intrinsic transverse momentum effects in a phenomenological way. We have investigated the role of different $J/\psi$ production mechanisms in NRQCD, focusing in particular on the CO contributions, both in the unpolarized cross sections and the transverse single-spin asymmetries.
It has been shown how the inclusion of transverse momentum effects can, effectively and efficiently, regulate the infrared divergences coming, within a collinear treatment, from CO contributions at low $P_T$. Two sets for the LDMEs, extracted in NRQCD but with quite different assumptions, have been considered. In both cases, a reasonably good description of unpolarized data up to $P_T\simeq 3$-4 GeV, as observed in $pp$ collisions at RHIC, has been reached within the theoretical uncertainties.

Estimates for the corresponding SSAs have been given as well, both for RHIC and LHCb kinematics, showing how $J/\psi$ production could be an invaluable tool to access the gluon Sivers function. Among the main results, we would like to emphasize the role of some CO contributions, namely those coming from the $2\to 1$ channels, which allow to be sensitive to the GSF even in the negative $x_F$ region. This could open a new potential strategy to get information on this important TMD.

We have then compared our predictions, obtained adopting a parametrization of the GSF as extracted from a previous fit to $A_N$ for inclusive pion production at mid-rapidities, against SSA data for $J/\psi$ production, showing a good agreement, independently of the quarkonium production mechanism. Indeed, very similar results are obtained both in the CSM and in NRQCD, and available data are not able to discriminate between these frameworks.

Further studies are certainly necessary concerning the production mechanism as well as the proper way to include TMD effects both in the calculation of unpolarized cross sections and SSAs. In this respect, the role of initial and final state interactions in the calculation of SSAs within a TMD approach could be extremely interesting. This issue will be addressed in a future work.

\section*{Acknowledgments}
This work is financially supported by Fondazione Sardegna under the project “Quarkonium at LHC energies”, CUP F71I17000160002 (University of Cagliari). This project has received funding from the European Union's Horizon 2020 research and innovation programme under grant agreement N.~824093.

%=======================================
%\bibliographystyle{apsrev}
%\bibliography{references}

\end{document}